\pdfoutput=1

 \documentclass[sigconf,authorversion,nonacm]{acmart}

\AtBeginDocument{%
  \providecommand\BibTeX{{%
    \normalfont B\kern-0.5em{\scshape i\kern-0.25em b}\kern-0.8em\TeX}}}

\setcopyright{acmcopyright}
\copyrightyear{2022}
\acmYear{2022}
\acmDOI{XXXXXXX.XXXXXXX}

\acmBooktitle{OO '00: ACM Symposium OO,
 ZZZ 00--00, 0000, OO, 00} 
\acmPrice{15.00}
\acmISBN{999-1-0000-XXXX-X/00/00}

\usepackage{hyperref, xcolor}
\hypersetup{%
  colorlinks=true,
  linkcolor=blue,
  linkbordercolor=red,
}

\makeatletter
\Hy@AtBeginDocument{%
  \def\@pdfborder{ 0 0 1 }
  \def\@pdfborderstyle{  /S/U/W 1.4}
}
\makeatother

\newcommand{\ie}{\textit{i}.\textit{e}.,}
\newcommand{\eg}{\textit{e}.\textit{g}.,}

\begin{document}

\title{SoK: How Not to Architect Your Next-Generation TEE Malware?}


\author{Kubilay Ahmet Küçük}
\authornote{Corresponding Author, reachable at kucuk@acm.org after the expiry of the institutional affiliations.}
\email{kucuk@cs.ox.ac.uk}
\orcid{0000-0001-5484-7096}
\affiliation{%
  \institution{ Cyber Security Centre, \\ University of Oxford, UK}
  \streetaddress{}
  \country{}
  \postcode{OX13PP}
}

\author{Steve Moyle}
\email{steve.moyle@cs.ox.ac.uk}
\orcid{0000-0003-4814-425X}
\affiliation{%
  \institution{ Cyber Security Centre, \\ University of Oxford, UK}
  \streetaddress{}
  \country{}
  \postcode{OX13PP}
}

\author{Andrew Martin}
\email{andrew.martin@cs.ox.ac.uk}
\orcid{0000-0002-8236-980X}
\affiliation{%
  \institution{ Cyber Security Centre, \\ University of Oxford, UK}
  \streetaddress{}
  \country{}
  \postcode{OX13PP}
}

\author{Alexandru Mereacre}
\email{mereacre@nquiringminds.com}
\orcid{0000-0002-5376-2194}
\affiliation{%
  \institution{Nquiringminds}
  \streetaddress{Southampton Science Park}
  \city{Southampton}
  \country{UK}}

\author{Nicholas Allott}
\email{nick@nquiringminds.com}
\orcid{0000-0001-7473-0565}
\affiliation{%
  \institution{Nquiringminds}
  \streetaddress{Southampton Science Park}
  \city{Southampton}
  \country{UK}}

\renewcommand{\shortauthors}{Küçük et al.}

\begin{abstract}
Besides Intel's SGX technology, there are long-running discussions on how trusted computing technologies can be used to cloak malware. 
Past research showed example methods of malicious activities utilising Flicker, Trusted Platform Module, and recently integrating with enclaves. 
We observe two ambiguous methodologies of malware development being associated with SGX, and it is crucial to systematise their details.
One methodology is to use the core SGX ecosystem to cloak malware; potentially affecting a large number of systems. 
The second methodology is to create a custom enclave not adhering to base assumptions of SGX, creating a demonstration code of malware behaviour with these incorrect assumptions; remaining local without any impact. 
We examine what malware aims to do in real-world scenarios and state-of-art techniques in malware evasion. 
We present multiple limitations of maintaining the SGX-assisted malware and evading it from anti-malware mechanisms.
The limitations make SGX enclaves a poor choice for achieving a successful malware campaign.
We systematise twelve misconceptions (myths) outlining how an overfit-malware using SGX weakens malware's existing abilities. 
We find the differences by comparing SGX assistance for malware with non-SGX malware (\ie~\emph{malware in the wild} in our paper).
We conclude that the use of hardware enclaves does not increase the preexisting attack surface, enables no new infection vector, and does not contribute any new methods to the stealthiness of malware.
\end{abstract}



\maketitle

\section*{Copyright}\small
Author's Manuscript. Open Access, Non-Commercial Version.\\
This manuscript is accepted on 20 September 2022 for publication.\\
Kubilay Ahmet Küçük, Steve Moyle, Andrew Martin, Alexandru Mereacre, and Nicholas Allott. 2022. SoK: How Not to Architect Your Next-Generation TEE Malware?. In Hardware and Architectural Support for Security and Privacy (HASP 22).

\normalsize
\section{Introduction}

Today it is highly likely that the majority of commodity systems are exposed to malicious software due to their highly complicated software stack.
One of the revolutionising techniques in fighting malware is to utilise trusted hardware enclaves.
However, using enclaves has also provoked discussions, and misconceptions leading to myths about whether attackers can utilise these enclaves for adversarial purposes.
Similar to end-user computers, home IoT deployments suffer from the same problem.
Nowadays, most home Internet of Things (IoT) deployments contain insecure, untrusted, outdated devices installed in houses, permanently spying on users.
We consider the malware living in commodity systems and home IoT deployments as the \emph{malware in the wild}.

An example of malware in the wild is a light bulb in a smart house.
The light bulb may contain a full Linux software stack, always online in the network, sniffing and spying on the home WiFi network, and constantly leaking user assets.
We group these types of harmful IoT devices as \emph{Evil Light Bulbs (ELB)}, as the malware living in a highly noisy commodity environment.
Trusted hardware technologies, \eg~Trusted Execution Environments (TEE), also fight back against ELB by securing the IoT deployments~\footnote{\url{https://www.iotsecurityfoundation.org/best-practice-guide-articles/device-secure-boot/}}.
We examine whether malware in ELB can utilise trusted hardware to be more powerful in attacks towards industrial IoT and smart home IoT deployments.
Our arguments and the misconceptions discussed in Section~\ref{sec:myths} for wild malware and enclave-based malware in computers are also valid for ELB in IoT deployments.

\textit{\textbf{Paper Structure: }} We structured this paper into four main categories.
First, we describe the frequently seen characteristics of an ideal malware.
Second, we demonstrate existing malware evasion techniques and high-scale, effortless malware distribution techniques.
We show non-TEE malware evasion techniques and a delivery campaign for two reasons; (1) to define the assumptions and review the scope of malware detection, (2) to demonstrate a real-world scenario on scalable infection to give readers an understanding of malware in the wild.
Third, we systematically evaluate twelve misconceptions on malware in TEE and present why these myths are far from the truth in practice.
Finally, we compare the malware in the wild with enclave-based malware and see if utilising enclaves provides any additional benefits to malware in practice.

\textit{\textbf{Theme: }} The theme of this paper is to evaluate TEE-assisted malware development from the perspective of malware developers (bad actors), understand malware characteristics and requirements, and evaluate systematically whether TEE assistance can fulfill the implementation details of a stronger malware.
Section~\ref{sec:limitationsMalwareSGX} relates to this theme by comparing a TEE assisted malware with wild malware in a system.
The relation of Section~\ref{sec:discussMalwareTEE} to the paper's theme is to discuss the remaining aspects potentially being abused by bad actors, and how these aspects should be considered in future trusted hardware designs.
We do not intend to discuss how attackers should be developing malware in next-generation TEE.
We show why attackers would find developing malware with TEE infeasible.
We hypothesise that malware in TEE gives no new advantages to attackers; we will discuss how it brings disadvantages in some cases.

\textit{\textbf{Contributions: }} This study presents the most comprehensive collection of misconceptions about malware in enclaves to date.
Our novel contribution is that we systematically show why malware becomes weaker due to trusted hardware.
We also clarify the ambiguity of \emph{malware in an SGX-enclave} and the \emph{malware in the core SGX ecosystem}.

\textit{\textbf{Concepts: }} Throughout the paper, we may use the following synonyms.
Wild malware refers to malware in commodity systems, malware in the untrusted world, ELB in IoT deployments, or malware in a highly noisy system.
Enclave-based malware refers to a malicious code placed in a custom developed enclave, SGX-malware, SGX-based malware, malware based on trusted hardware, or malware in TEE.
In order to distinguish the core SGX ecosystem from custom developed enclaves, we provide a detailed definition in Section~\ref{sec:ecosystem}.

\subsection{Scope of Malware}

Malware presence can be in various forms; \emph{virus, worm, trojan, zombie/botnet, phishing/spam, spyware, keylogger, sniffer} and \emph{ransomware}.
Throughout this paper, we refer to this section for the precise scope of the \emph{malware}. 
We take into account specifically the malicious software targeting Windows PCs.
Nevertheless, most of our systematisation can also be valid for Linux distributions and OS X systems (Intel, non-ARM).
Malware targeting other commodity operating systems, advanced persistent threats (APT), and other relatively harmful software can be explored outside of this paper. 
We keep relatively badly behaving software outside of the malware definition.
A bad program is one, when found and analysed by an AV company, is subsequently detected by their malware detection systems.
Finally, we consider malware as malicious software tailored for disruption, damage and unauthorised access to the commodity computer system.
We describe the characteristics of malware in more detail in Section~\ref{sec:characteristics}.
Similar to malware in a single computer, an IoT deployment with tens of devices in a house suffers from the same problems due to ELB.

\subsection{\textbf{Characteristics of Malware}}
\label{sec:characteristics}
The ideal malware must~\footnote{Others may structure these characteristics or malware patterns in an alternative way.} have the characteristics listed below for a successful malware campaign. 
However, one of the fundamental issues is that these characteristics can also be seen in benign and legitimate software. 
Having these characteristics in benign software makes malware detection a challenging task for anti-malware mechanisms.

\subsubsection{\textbf{Persistence across boot cycles/load at startup}}
\label{sec:persistence}
The malware must be able to reload itself whenever the system is available. 
This can be done via scheduled events, registry entries, drivers and functions in other processes.
Section~\ref{sec:mythE} presents the persistence aspect of enclave-assisted malware.

\subsubsection{\textbf{Communication on demand}}
\label{sec:communication}
Ideally, the malware must be able to communicate with the author to receive commands or leak information whenever required. 
Communication might be necessary for reselling the network of victims, fetching a new payload or ready-to-use new exploits, or equipping malware with up-to-date zero days to be exploited.
A secure communication channel is also required for the key management; whether the keys are managed locally or through external servers, the malware must have a direct or indirect key management channel.
We present the communication aspects in Section~\ref{sec:mythC} and Section~\ref{sec:mythF}, and key management aspects in Section~\ref{sec:mythB} and Section~\ref{sec:mythD}.

\subsubsection{\textbf{Full Un-Detection (FUD)}}
\label{sec:detection}
The malware must remain fully undetected throughout its life cycle against any possible detection mechanisms.
At a minimum, this can be done via resilience to all anti-virus software products in the market. 
FUD checks can be done offline via automated environments without leaking the know-how of malware.
We present the arguments on detection in Section~\ref{sec:mythA} and Section~\ref{sec:mythG}.

\subsubsection{\textbf{Access to system calls/APIs}}
\label{sec:sysCall}
Ideal malware must be able to access kernel functionalities, system calls, and Application Programming Interfaces (APIs) present for low-level operations.
Access to the full system is especially necessary for targeted attacks where a victim user's core assets are extracted.
Section~\ref{sec:mythH} presents the arguments on syscalls.

\subsubsection{\textbf{The highest possible privileges}}
\label{sec:priviliges}
The malware must escalate into privilege levels present in a system as high as possible.
Most attacks may begin at the lowest level of permissions; however, successful malware must be equipped with the necessary payloads to exploit the rest of the system.
Section~\ref{sec:mythE} and Section~\ref{sec:mythI} present the arguments on malware privileges.

\subsubsection{\textbf{Unrestricted resources and assets}}
\label{sec:resources}
Ideally, malware must be able to use all system resources without restrictions.
Resources may refer to all devices/components other than computational abilities and full memory access.
These resources may also contain the target user assets and valuable information.
We present the arguments on resources in Section~\ref{sec:mythJ} and on user assets in Section~\ref{sec:mythD}.

\subsubsection{\textbf{Infect more targets, audience and availability}}
\label{sec:audience}
Ideally, malware must aim to infect more victims in a campaign. 
Infection rate can be a success measure for malware.
A malicious attack targeting one specific system might fall into the APT category.
For compatibility, the target system must fulfil the requirements of the malware at infection time.
Further, if the malware has specific dependencies to operate, these dependencies must continue to be available.
We present the arguments on target victims in Section~\ref{sec:mythK} and malware availability in Section~\ref{sec:mythC}.

\subsubsection{\textbf{Maximise revenue/profit and long-term maintenance}}
\label{sec:maintenance}
The goal of malware should be to increase its profit above all.
The monetisation of a campaign might be a way to see the profit, but the overall damage or the information extracted can also be of profit to the malware authors.
The malware would require maintenance for long terms jobs or while reselling or in case of an ownership transfer.
Section~\ref{sec:mythF} and Section~\ref{sec:mythL} present arguments on malware maintenance.

\subsection{Focused TEE: The SGX Ecosystem}
\label{sec:ecosystem}
TEE provide hardware-assisted new security capabilities as specified in Global Platform documentations~\footnote{\url{https://globalplatform.org/technical-committees/trusted-execution-environment-tee-committee/}}.
We focus on Intel's Software Guard eXtensions (SGX), widely available for commodity systems today in the market.
SGX-enabled hardware products allow applications to be executed in a protected memory region called an enclave.
Enclaves aim to protect sensitive workloads, process secret information and provide strong isolation for applications.
To examine better whether enclaves can fundamentally help to cloak malware in contrast to its design goals, we briefly describe the SGX ecosystem.
SGX was introduced in new hardware instructions (host + user level\footnote{13 host + 5 user; it depends on if we count actual instructions or SGX leaf instructions.}, and VMM instructions added), and is a hardware-level feature, implemented as xucode\footnote{Online. \url{https://www.intel.com/content/www/us/en/developer/articles/technical/software-security-guidance/secure-coding/xucode-implementing-complex-instruction-flows.html}} in the CPU package.
The implementation can be considered more or less as executing on the ring level minus three (-3), and SGX can be configured in the firmware. 
The memory amount allocated for SGX is predefined and known at the system start.
The amount of encrypted memory pages that applications can benefit from depends on the enclave's requirements at run time.

Intel provides a native Software Development Kit (SDK) for enclave development, which contains trusted libraries (nowadays absolutely required in any enclave) to be imported into enclaves.
There are also other SDKs for enclave development, but we keep them out of the scope of the core SGX ecosystem.
For example, Fortranix Enclave Development Platform (EDP, \url{https://edp.fortanix.com/}), is one of the most advanced environments to develop secure enclaves, utilising RUST language.
Enclave developers are responsible for keeping their development environment malware-free, and only trusted libraries and trusted SDKs must be utilised.
Intel's SGX SDK provides privileged architectural enclaves such as Launch Enclave, Quoting Enclave and Platform Service Enclave (LE, QE, PSE) and others.
These are assistive enclaves for the custom enclaves developed; they are built with Intel's SGX SDK.
Enclaves can have manufacturing, provisioning and attestation life cycles in their production and execution periods.
Finally, there are crypto libraries and protocols (\eg~for communication) utilised in enclaves.
These points define the SGX ecosystem, where SGX itself can be held responsible for a bad design choice.
After all, the custom enclave code is outside of the SGX ecosystem, and outside of a TEE ecosystem.
If the custom enclave contains a vulnerable piece of software code programmed by the enclave developer, the security guarantees offered by the hardware may be diminished, and SGX instructions should not be blamed for such cases.
The composition of the underlying hardware and the developed software plays a crucial role in secure application development.
Some of the important questions:

\begin{itemize}
\item Is persistent-malware taking advantage of any of these features in the core SGX ecosystem? 
\item Does malware gain superior features or become weaker by utilising these features? 
\item Where is the actual location of malware (\eg~in the microcode, or in developer-defined enclave-code)?
\end{itemize}

\subsection{Related Work}
\label{sec:relatedWork}
There are example studies in the literature~\cite{chen2019sgxpectre,kocher2019spectre,lipp2018meltdown,van2018foreshadow,schwarz2017malware} for bad CPU design choices, micro-architectural attacks and the known-to-be-vulnerable software chunks used in enclaves.
When CPU design choices cause a security issue, microcode can receive a trusted computing base (TCB) update~\cite{intel_microcodeUpdate}. 
In fact, the system's microcode gets patched at each boot cycle.
If an attack requires full kernel and OS control, this can be mitigated by a measured boot of a formally verified kernel.
The applications containing vulnerable software must update their enclave code base and revoke any execution permissions for the older versions.

We examine the question of whether developing a fine-grained malware by utilising TEE makes it stronger or weaker~\cite{itl_SGX_rutkovska1,itl_SGX_rutkovska2, Marschalek_wolf_sgx, schwarz2019practical}. 
There are long-lasting discussions based on these questions: 

\begin{itemize}
\item Can SGX help to deliver malware payloads?
\item Can TEE help ransomware (\ie~for operations of data copy/ encryption/ delete, key generation/ storage, persistence, communication)?
\item What secret operations can an enclave handle/execute inside an isolated memory region? 
\end{itemize}

The purpose of enclaves is to allow secret operations; nevertheless, any secret operations with malicious intentions are similar to independent, remote operations.
We shall further discuss these questions in detail in the scope of this paper.

Enterprise-level anti-malware solutions with advanced memory protections can apply to enclave-based malicious software.
Memory Exploit Mitigation (MEM) from Symantec~\cite{symantec_sgxTamas_Rudnai}, and other defensive techniques as Data Execution Prevention (DEP), Address Space Layout Randomization (ASLR), Structured Exception Handling Overwrite Protection (SEHOP) and Return-oriented Programming (ROP) protections are effective for detection and mitigation purposes~\cite{symantec_broadcomEndpoint}.

\section{Existing Difficulties in Malware Detection}
Malware detection in a high entropy system has been a difficult challenge for decades.
The complexity of an end-user's commodity system and the similar patterns of benign and malicious software make malware detection more difficult.
We show example methods of non-SGX malware for malware stealthiness and malware delivery (\ie~malware without enclaves, malware without trusted execution environments) in Section~\ref{sec:case1} and Section~\ref{sec:case2}.
These methods continue to achieve successful malware campaigns.
Throughout our discussions, we examine whether an enclave-assistance can make a malware campaign more successful, or result in its quicker failure.
We explain the misconceptions in Section~\ref{sec:myths}.
The reason why we show a non-SGX malware is to be able to compare it with a potential SGX-based malware systematically and draw a clear picture of their comparison of whether the existing capabilities increase or decrease.


\subsection{Case Study: (Non-SGX) Malware Infection in Memory}
\label{sec:case1}
Malware in a commodity system (\eg~in Windows PC) can escalate its privileges to have full memory access.
We present three common techniques for a malicious payload to spread and continue to execute with benign processes.
Depending on how sophisticated implementation a malware has, detection might be very difficult.
Most of these actions can be seen in legitimate software as well.
Thus the malicious behaviour can often bypass the detection mechanisms.

Software-based memory encryption and obfuscation techniques can hide malware in run-time and make reverse engineering difficult.
We show three of these commonly used techniques.
Due to real-time mutations, \eg~malware polymorphism, old-time static code analysis and detection techniques (signature-based) can fail.
Failure of static analysis techniques and the rise of dynamic malware identity motivated the industry to bring promising behaviour-based dynamic analysis techniques, even if the memory is encrypted or obfuscated.
Today, anti-malware companies use advanced memory protection and analysis solutions explained in Section~\ref{sec:relatedWork} to fight back the memory-based evasion techniques.
We highlight the following techniques to comprehend the myths in Section~\ref{sec:myths}; SGX run-time memory encryption does not give the anti-malware industry a new challenge.

This paper assumes that users (implementing or using anti-malware solutions) will not opt for old-time static code analysis or signature-based detection.
If they rely on signatures only (old-fashioned), some readers may think that enclaves can hide the code with memory encryption.
The code loaded into an enclave is initially a static binary and is open for inspection or reverse engineering.
The potential threat with enclaves might come at runtime where an enclave fetches dynamic content (potentially malicious); this is where we argue that the threat vector is not new for the anti-malware industry.
Section~\ref{sec:mythA} on memory encryption and Section~\ref{sec:mythJ} on memory access are related to this section.

\textbf{RunPE:}
Windows computers utilise portable executable (PE) format.
In the RunPE technique, the malware replaces the memory content of a legitimate process (\ie~belonging to a benign system service) with the malicious payload.
The payload does not need to be extracted all at once; malware can extract small pieces of malicious content depending on its goals and revert back to the original memory.
Linux systems use Executable and Linkable Format (ELF); a similar technique can be called RunELF, where the content of a legitimate process is replaced with a malicious payload at runtime.

\textbf{PE/ELF injection:}
Malware (\eg~a C binary) in a commodity system with low-level privileges can access the full content of the memory.
Instead of replacing the process memory, malware can also allocate a new memory region, inject its payload (PE or ELF) and align/compute the memory addresses so that the execution can continue.
Similar to the RunPE method, the payload extraction happens at runtime and does not utilise the persistent storage.

\textbf{DLL/SO injection:}
Alternatively, malware can prepare a modified library, Dynamically Linked Library (DLL) or Shared Object (SO), containing the malicious functions ready in the persistent storage to be called.
Often in commodity systems, even if the system had measured boot, the libraries in the disk can be replaced by different versions highly likely containing malicious payloads.

\subsection{Case Study: (Non-SGX) Drive-by Malware Distribution}
\label{sec:case2}
Besides the memory operations for stealthiness, a malware can be capable of escalating to higher privileges.
It is crucial to examine how it can spread in the first instance and escalate its privileges.
We implemented a scalable malware delivery campaign for this case study and reported it with responsible disclosure steps.
Note that it is no longer possible to follow the exact steps as (1) Facebook does not allow the inclusion of third-party video players anymore, and (2) Adobe no longer supports Flash Player since the end of 2020~\footnote{Although flash technology is no longer available, we anticipate that a similar set of issues can be seen with WebAssembly applications and other client-side containers.}.
No real accounts were used in the proof-of-concept (PoC) experiment, only the test accounts generated by Facebook for bug bounty purposes were used.

The following PoC is an example of the malware in wild; it gives us tangible facts to establish a comparison with potential TEE-assisted malware. 
We discuss more details of the enclave-assisted malware and the malware in wild in Section~\ref{sec:comparison}.
Section~\ref{sec:mythC} on malware delivery and Section~\ref{sec:mythK} on target victims are related to this experiment.

\textbf{Scalable drive-by malware delivery campaign:}
At the time of our experiment, Facebook was allowing the embedding of externally hosted ShockWave Flash (SWF) video players on users' walls.
We developed a custom video player in ActionScript 3.0, streaming legitimate content, hosted on our HTTPS-enabled external web server. 
Having a TLS connection on the host server was the only condition by Facebook to cache the custom video player and stream the videos on users' walls inline.
Once the source link has been distributed through a high number of shares (\eg~to millions of users), we updated the initially benign video player containing \emph{known} exploits (for OS X, Linux, Windows) available in the public space~\footnote{Known CVEs, existing exploit where the patch is available. The development details of the exploit itself are out of scope for our paper.}.
Although it is expected for most of the target users to have up-to-date (non-vulnerable) versions of flash and sand-boxed browsers, there are a sufficient number of users with vulnerable software.

We demonstrated a proof-of-concept video to the Facebook security team, triggering other existing vulnerabilities in the system (\eg~enabling a drive-by Java downloader, triggering a local media player's vulnerabilities, calling system applications and accessing kernel functions).
To infect a system~\footnote{A similar simplicity of infection can also be seen in an IoT deployment; ELB can cause infection once the lights are turned on.}, it is sufficient for victim users to watch a video on their Facebook feed.
Initially, the Facebook security team responded that (1) the nature of the platform is to allow third-party content to be circulated (2) their internal whitelisting/blacklisting mechanism can block the source link as soon as it is detected.
Later, (and currently) the functionality of embedding third-party video players is completely removed.
Crucially, we did not need any unknown zero-day vulnerabilities.
As Facebook provided the highly scalable distribution network, we were able to demonstrate this attack with known exploits that were not patched by all users. 
Having known exploit payloads in SWF video players also proved to us that Facebook's internal scanning mechanisms before caching any external resource were inadequate.
Finally, although the inline video embedding/playing feature is removed, the danger of distributing malware in social media continues. 
It is always possible to change the content of an external link once it has been shared millions of times.
The worst part is that media contents can be changed depending on user location and personal views (\eg~political), giving an extended ground for manipulation.
Once a user's computer is infected, all of the content in a social media account can be tampered with locally at the client side through malware.
Associating users' true identities with infected machines gives attackers a highly valuable asset in malware reselling.

\section{Misconceptions about enclave assisted malware}
\label{sec:myths}
Although the public launch of SGX technology was in 2015, SGX-assisted malware discussions within the research and security community go back to 2013~\cite{itl_SGX_rutkovska1,itl_SGX_rutkovska2} at least\footnote{Researchers may be able to trace TEE-based malware discussions considering older technologies with DRTM, TXT or similar.}.
Our goal in this section is to stretch these claims to study and understand them correctly; this may result in some of the myths being read as hand-waving.
We aim to give readers an understanding of how dangerous SGX-assisted malware can be and how future TEE designers can benefit from our systematisation in order to avoid TEE features being abused by bad actors to strengthen their malware.
Besides our work, Symantec has also published~\cite{symantec_sgxTamas_Rudnai} a written statement that claims behind SGX malware may not be as bad as it is \emph{believed}, and they may remain as myths only.
Here in our systematisation, we present them as myths or misconceptions interchangeably.
In accumulation and conciseness of these claims, we benefited from past SGX Community Day discussions and presentations~\cite{kucuk_malware2020}.

The structure we create may serve as a template for future discussions, and may evolve with support or criticism from the community.
In this section, we structure~\footnote{Through SGX research community days, community's written statements, and other references found in this document.} the claims about malware utilising enclaves especially crafted for SGX-enabled trusted hardware.
Similar characteristics and claims are valid for ELB utilising trusted hardware capabilities to attack IoT deployments internally.
The claims and arguments behind these myths are derived from ITL resources~\cite{itl_SGX_rutkovska1, itl_SGX_rutkovska2}, industry/practitioners conferences \footnote{\url{https://www.zdnet.com/article/researchers-hide-malware-in-intel-sgx-enclaves/}}~\cite{schwarz2019practical}, industrial vendor resources~\cite{symantec_sgxTamas_Rudnai}, and  are compiled from Intel's SGX Community Days discussions and presentations~\cite{kucuk_malware2020}. 

\subsection{\textbf{Myth: Enclave's memory encryption engine will hide the malware}}
\label{sec:mythA}
\textit{Relates to the FUD characteristic of malware in Section~\ref{sec:detection}.
Section~\ref{sec:mythG} presents further arguments on malware detection.}

\textbf{Supporting Arguments: }
SGX provides a memory encryption engine for trusted applications.
The content of the memory pages of enclaves is not visible at runtime.
For example, the memory of enclaves cannot be inspected directly.
Comparing it to an untrusted part of the memory, enclaves can help to hide the malware (\eg~malicious payload or malicious behaviour).

\textbf{Counter Arguments: }
Enclave binaries are inspectable in the disk, and anyone can dump the initial content of the enclave, as shown in~\cite{kuccuk2019managing}.
The memory locations of enclaves are the most visible part of the system.
The network traffic into the enclaves is visible; although traffic is encrypted, most of the network analysis techniques still apply.
All of the disk operations, input and output operations on main memory, the interface operations with the untrusted part, usage patterns and CPU utilisation of the enclaves are visible.
These points make enclave-based malware even more visible than malware in the wild.
Hardware-assisted encryption instead of software-based encryption may give stronger security guarantees for malware; nevertheless, the actual detection of malware is only possible through the behavioural patterns.
In other words, detection mechanisms do not necessarily require decryption of the malware.
Tracking malware in a complex and noisy environment is more difficult than tracking the behaviour of an enclave which is entirely dependent on resources managed by the OS.

\textbf{Reference and Rebuttal: }
\cite{van2016design} states that enclave can hide the malware, they conclude that malware can benefit from SGX.
In fact, they only download static plain-text data into an enclave, with no proof of execution or malicious activity where SGX directly contributes.
They ignore multiple facts how the initial enclave was launched, where did this enclave connect or who supervised it.

\subsection{\textbf{Myth: Enclaves will generate encryption keys for each malware payload}}
\label{sec:mythB}
\textit{These arguments relate to malware key management requirements in Section~\ref{sec:communication}. 
We discuss the scalability and ransomware aspects of key management in Section~\ref{sec:mythD}.
}

\textbf{Supporting Arguments: }
Enclaves can generate private keys inside encrypted memory regions.
These keys never leave the private memory.
This can help malware to maintain its key generation and key storage problems.
Further, malware can generate unique keys for each victim.

\textbf{Counter Arguments: }
The instruction used for key generation \texttt{EGETKEY} is bound to the enclave binary identity.
Enclave is measured at load time once, and any payload that the enclave fetches later does not change the enclave identity.
The source code of the initial binary is inspectable on the disk.
Enclave ID, therefore, enclave-based derived keys are based on initial binary (object code can be dumped).
Generated sealing keys can be derived from initial enclave measurement and can be derived from root sealing key as well.
An attacker can produce unique keys for the victims, bound to hardware and software.
But, generating independent keys in an enclave is similar to generating them at a remote location.
The key point is that the dynamically fetched payload is not part of the key generation through SGX, it is done through already existing manual efforts.
Overall, this does not provide the malware author with anything new or any superior feature.
The main enclave identity remains the same as the initial binary.
Nevertheless, SGX's newer Key Sharing and Separation (KSS) feature must be examined separately.
Although KSS does not change the enclave ID for attestation purposes, optionally, KSS can include the identity of the newly fetched payload for sealing key generation.
This may give a unique key based on the payload.

\textbf{Reference and Rebuttal: } 
\cite{bhudia2021ransomclave} shows that attackers can generate public-private key pair inside an enclave and attackers can use them for encrypting user files with the public key outside of the enclave. 
The paper makes many assumptions far from reality that all conditions must be set for a successful attack.
Further, it tries to solve a problem that does not exist anymore; ransomware gangs changed how they carry out their attacks.
The use of an enclave for key generation makes the job of attackers more difficult.

\subsection{\textbf{Myth: Enclaves will secretly deliver malware}}
\label{sec:mythC}
\textit{These arguments relate to communication aspects in Section~\ref{sec:communication} and availability requirements in Section~\ref{sec:audience}. 
Section~\ref{sec:mythF} presents communication of malware towards establishing secure channels.
}

\textbf{Supporting Arguments: }
Enclaves can fetch secret payloads into the private memory regions at runtime.
This may allow enclaves to be a point of malware distribution.
Enclaves might be used as stateless malware carrying embassies.

\textbf{Counter Arguments: }
Enclaves do not receive the payloads out of anywhere arbitrary. 
The network patterns and their behaviour can be observed.
Further, enclaves are not always available in a system.
An enclave's responsiveness (\ie~opportunity to real-time respond to an attacker) is relatively low compared to the system's other services.
If an enclave allows malicious payload at runtime, it must utilise public key infrastructure (PKI) to ensure that only the attacker can fetch a payload.
Otherwise, malware reverse engineers can also inject healing code inside an enclave and extract its private keys.
Overall, attackers utilise classical PKI techniques for malware delivery as usual, and do not profit directly from SGX's features.
~\cite{DBLP:journals/corr/abs-2104-03868} explains the past use of PKI, for potentially malicious or benign purposes.
SGX enclaves are just another point they might choose for malware delivery, but their low availability makes them a bad choice for high-scale infection campaigns.
Launch Enclave (LE) is a special enclave to account for malware in an enclave. 
It can be used to determine whether the enclave may be launched on the platform, utilising launch tokens and policies.

\textbf{Reference and Rebuttal: } 
\cite{van2016design} states that stalling mechanism can delay the malware behaviour and help to delivery.
Time measurement inside an enclave is difficult, they develop a mechanism to measure the time, however, the paper does not show any solid proofs for malware delivery.
They ignore the actual infection point in practice and jump to the behavioural detection discussion.

\subsection{\textbf{Myth: Enclaves will scale and ease ransomware key management}}
\label{sec:mythD}
The following arguments relate to key management in Section~\ref{sec:communication} and access to user assets in Section~\ref{sec:resources}.

\textbf{Supporting Arguments: }
Ransomware key management (generating and storing the keys, or regenerating the keys at a later stage) can be easily achieved within enclaves.
Every victim can have a unique key based on their machine identity.
Similar to any digital rights management software designed for enclaves, ransomware can utilise the same features.

\textbf{Counter Arguments: }
Using enclaves for a ransomware leaves more behaviour patterns in a system.
Ransomware must still copy the data into enclaves if files will be used for encryption purposes.
If the enclave generates the keys and gives the secret keys to the untrusted world for encryption, this would be considered a design failure for the ransomware.
Either the encryption can take place outside the enclave with the released public key or inside the enclave.
Outside encryption is susceptible to existing detection mechanisms.
Technically, carrying all of the user's data into the malicious enclave is not a feasible operation.
The reason is; data transfer is dependent on outside operations requiring multiple assumptions.
Furthermore, enclave-based key management adds an extra step to the current ransomware key management mechanisms.
The healthiest way to maintain ransomware keys still remains as remote resources.
Additionally, a ransomware must encrypt a considerable amount of data over time without being detected, making the existing noise level of the untrusted world essentially a better choice for malware authors.

\textbf{Reference and Rebuttal: }
\cite{bhudia2021ransomclave}'s threat model starts with the assumptions that malware delivery and installations are already done, system has no anti-virus and no anti-ransomware mechanisms, they only consider key capturing mechanisms~\cite{kolodenker2017paybreak} which attackers already solved since five years.
They argue that block-chain based mechanisms will automate the ransom payments.
In overall, their ransomware is supported more by their assumptions and block-chain based methods, there is no evidence that use of enclaves resulted a stronger ransomware.

\subsection{\textbf{Myth: Enclave-assisted malware will be persistent in the system}}
\label{sec:mythE}
\textit{These arguments relate to persistence in Section~\ref{sec:persistence} and malware privileges in Section~\ref{sec:priviliges}.
Section~\ref{sec:mythI} presents the arguments on privileges at the custom enclave code (ring 3) and privileges of the micro-code level (ring -3). }

\textbf{Supporting Arguments: }
SGX-enabled machines can start-up an enclave whenever it is triggered.
This can allow private operations to be completed on demand.
Malware can live in a system as persistent through enclaves.
Lower-level hardware assistance can enable persistent rootkits in a system.

\textbf{Counter Arguments: }
Enclaves themselves are not persistent pieces of software applications.
They have a life cycle of \texttt{ECREATE} where the enclave is created and \texttt{EDESTROY} to kill the enclave.
Such operations were  explained and measured in~\cite{kuccuk2016exploring}.
Enclaves operate at the user level (ring 3) and although their execution is supported by the low-level hardware features, they do not operate at any lower privilege levels in the system.
Enclave-based malware cannot be persistent on its own.
Outside system support for persistence would create an extra burden.
If a malware can be persistent in the untrusted world, trying to utilise an enclave may only make it less persistent.

\textbf{Reference and Rebuttal: } 
\cite{toffalini2021snakegx} states that their malicious operations can reach persistence through SGX enclaves.
They assume custom enclave code to be vulnerable, not exploiting anything in the core SGX ecosystem. 
They claim persistence, but it is completely dependent on user-level host enclave, in reality, this offers a very weak persistence.

\subsection{\textbf{Myth: The malware inside an enclave will communicate independently}}
\label{sec:mythF}
\textit{These arguments relate to the communication requirement in Section~\ref{sec:communication} and the malware control in Section~\ref{sec:maintenance}.
Section~\ref{sec:mythL} presents further arguments on malware maintenance with ownership transfer.}

\textbf{Supporting Arguments: }
Malicious code executing inside a private memory can communicate with the outside world through HTTP calls and secure channels.
This can help to leak information from the user's system.

\textbf{Counter Arguments: }
For communication, enclaves rely on untrusted channels.
Although enclave-based malware can send arbitrary data to the outside world, the classical method applies before malware fetches or copies any private data from the disk or memory.
This is not a new issue; most of the applications in a commodity computer can steal users' private files at any moment.
Enclaves, in fact, have more dependencies for their communication with the outside world.
The technical dependencies and libraries an enclave minimally may need is presented in~\cite{kuccuk2016exploring}. 
An alternative but larger option can be library OSes,~\eg with interpreter enclaves as explained in~\cite{DBLP:journals/corr/abs-2104-03868}.
They can also leave clear traces of their communication pattern.
Monitoring the whole system's (host) communication is more demanding than monitoring an enclave.

\textbf{Reference and Rebuttal: } 
\cite{van2016design} performs attestation with a remote server and fetch a payload from outside.
They ignore the fact that all communication channels were dependent on untrusted channels where in reality they would not be able to freely communicate with any malicious server.

\subsection{\textbf{Myth: TEE-based malware will be FUD}}
\label{sec:mythG}
\textit{These arguments relate to malware evasion in Section~\ref{sec:detection}.}

\textbf{Supporting Arguments: }
Antivirus software products cannot directly perform a scan on the enclaves' memory.
Enclaves can keep malicious payloads away from scanners.
This can cause a shortcut in achieving FUD malware.

\textbf{Counter Arguments: }
An encrypted piece of the non-executing payload does not harm the system.
Enclave software or non-enclave software can both have pre-encrypted payloads.
As soon as a piece of code starts to execute, the same malware detection mechanisms apply.
A malware inside an enclave must extract instructions to an untrusted world or  perform operations in the main memory in order to reach valuable user assets (or towards its campaign goals).
Existing research and the techniques on malware detection apply for malware behaviour in the untrusted world.

\textbf{Reference and Rebuttal: } 
\cite{schwarz2020malware} states that their malware in an SGX enclave remains completely/entirely invisible to anti-virus software and even to ring 0.
The argument is ambigious and expected as their attack does not leave the SGX ecosystem nor enclaves, it crosses the boundaries of enclave isolation through caches.
The attack can be successful only with the specific requirements and assumptions.
In reality, being invisible to the antivirus mechanisms means that the malware remains invisible while being actively inspected and being in operation against the full system resources or assets.
If their malware attacked to the actual system, it would be then visible; their invisibility is relying upon having no attack to the kernel or similar components, hence the kernel does not see anything.

\subsection{\textbf{Myth: SGX based malware will access System APIs}}
\label{sec:mythH}
\textit{These arguments relate to system calls capabilities in Section~\ref{sec:sysCall}.}

\textbf{Supporting Arguments: }
Malware can reach system calls, WinAPIs and kernel functionalities from the enclave.
Accessing ring 0 level resources can boost the capabilities of the malware.
Within the enclave memory, malware can remain hidden and still perform a scan for vulnerabilities in the system to reach the kernel.

\textbf{Counter Arguments: }
Enclaves operate at user level privileges (ring 3).
In order to escalate into higher privileges or to access any system calls, the malware must go through the untrusted parts of applications, or perform operations on the main memory~\cite{SGXEnclaveWritersGuide}.
The malicious behaviour from the enclave towards the system can be monitored and restricted.
Depending on the programming model and the software development kit, the enclave may have no access to the system at all. 
The enclave may be given a Library Operating System (LibOS) to utilise dependencies internally without needing to make calls to the outside system.

\textbf{Reference: } 
\cite{symantec_sgxTamas_Rudnai} answers the questions on what SGX malware may need to do in order to access system APIs.
The arguments remain valid.

\subsection{\textbf{Myth: Malware will have the highest privileges through SGX}}
\label{sec:mythI}
\textit{These arguments relate to malware privileges in Section~\ref{sec:priviliges}.}

\textbf{Supporting Arguments: }
SGX hardware is a microcode implementation and has the highest privileges in a system.
A malware utilising enclaves can benefit from these high privileges.
If malware operates at ring level minus three, no detection mechanism can catch its malicious activities.

\textbf{Counter Arguments: }
Having a malware inside the core SGX ecosystem is not the same as having a malware in a custom developed enclave.
Enclaves operate at the user level, and their abilities are supported by the extended CPU instructions.
If a malware exploits CPU-level bugs on SGX, it might have the highest privileges.
Otherwise, malware in an enclave can only benefit from the user-level privileges.

\subsection{\textbf{Myth: Enclave assistance will give malware full memory access}}
\label{sec:mythJ}
\textit{These arguments relate to malware resources in Section~\ref{sec:resources}.}

\textbf{Supporting Arguments: }
Untrusted applications cannot see the memory of trusted enclave applications.
But enclaves can see the memory of the all main memory content.
This gives a malware inside an enclave a superior power.
\textbf{Counter Arguments: }
Untrusted applications can already see the full main memory content.
The ability of enclaves to see the full memory is not a new feature on top of any other usual application.
If an enclave performs operations on the main memory, the input/output traffic is visible to the operating system.

\subsection{\textbf{Myth: TEE will help malware to target more victims}}
\label{sec:mythK}
\textit{These arguments relate to malware delivery in Section~\ref{sec:audience}.}

\textbf{Supporting Arguments: }
SGX-enabled hardware products are standardised in Intel CPUs.
Since the end of 2015, most of the new laptops, desktops, and other machines have SGX bit in their firmware.
Users may not be well informed about SGX features and it might be left default enabled.
The applications utilising enclaves can also drop malware in users' machines.
Having SGX enabled computers can increase the number of victims infected by malware.
\textbf{Counter Arguments: }
From a malware author's point of view, non-SGX malware can target more users than a malware tailored for SGX.
A malware targeting the general audience (both SGX-enabled and non-SGX machines) can infect a higher number of victims.
SGX may be unavailable at any moment; therefore, relying on SGX being enabled is a poor design choice for malware.
The host can kill an enclave with suspicious behaviour (executing for an unnecessarily long time, memory operations, suspicious network traffic) from outside (other than the internal destruction function).
If a malware requires an enclave to be restarted in order to be in operation, this will make it a very ineffective malware.
The malware must be independent as much as possible; utilising enclaves give malware a sizeable hump to carry at all times.
Intel SGX is no longer available in consumer devices, unfortunately.
Only servers have Intel SGX at the moment.

\subsection{\textbf{Myth: Malware inside an enclave is easier to maintain}}
\label{sec:mythL}
\textit{These arguments relate to malware maintenance in Section~\ref{sec:maintenance}.}

\textbf{Supporting Arguments: }
Enclaves aim to minimise the trusted computing base. 
They contain a relatively small amount of code to make formal verification feasible.
Enclaves can fetch a new malware payload whenever necessary and act as an update mechanism for malware.
This can make malware maintenance cheap and more sustainable.
\textbf{Counter Arguments: }
Enclave development has many dependencies and it is an evolving ecosystem\footnote{\url{https://download.01.org/intel-sgx/latest/linux-latest/docs/Intel_SGX_SDK_Release_Notes_Linux_2.16_Open_Source.pdf}}.
A malware designed for enclaves would actually require additional engineering work to be always compatible.
Additional compatibility issues can cause malware to fail to operate in a high number of systems.
Crafting a malware for SGX brings additional development costs, maintenance requirements, and lowers the number of target machines to infect or continue to operate on.

\section{The Limitations of SGX-Malware}
\label{sec:limitationsMalwareSGX}

An SGX-based malware is not impossible to produce, but in practice, such malware would become weaker as explained in Section~\ref{sec:myths}.
Further, we now compare SGX-based malware with a malware in the wild for their characteristics listed in Section~\ref{sec:characteristics}.
Now in this section, we present systematic limitations (or disadvantages\footnote{All malware has disadvantages to anyone unfortunate to be infected with it, but we refer here to disadvantages for a successful malware campaign carried out by malware authors.} for the malware developers/bad actors) of utilising a TEE for developing malware.
ELB in IoT deployment can deploy attacks against user assets similar to wild malware in commodity computers.
We argue that using TEE in the development of malicious IoT devices (in the domain of ELB) will restrict their capabilities for the reasons listed in the following section.

\begin{table*}
  \caption{Malware in the wild (untrusted high noise system, non-SGX) in comparison to malware in an enclave utilising SGX features.  $\oplus$ Enclave enhances/strengthens the malware. $\oslash$ Enclave has no impact.  $\ominus$ Enclave weakens the malware. }
    \centering
    \begin{tabular}{ | l | l | l | l | }
    \hline
\textbf{Characteristics in Misconceptions}   &  \textbf{Malware in Enclave (ME)}    &  \textbf{Malware/ELB in wild}                 & \textbf{Conclusion} ($\oplus$, $\oslash$, $\ominus$): and Reason  \\    \hline
 \ref{sec:persistence} \textbf{Persistence} in \ref{sec:mythE}      & Needs  to trigger               & Services, drivers           &   $\ominus$: ME is less persistent.   \\   \hline
\ref{sec:communication} \textbf{Communication} in \ref{sec:mythC}, \ref{sec:mythF}   & Needs  untrusted channels & In high noise    &   $\ominus$: ME has longer path to communicate.    \\  
  \textbf{Key management} in  \ref{sec:mythB}  \ref{sec:mythD}     & Extra steps for payload      &  Existing  PKI                  &   $\ominus$: ME has more crypto operations.   \\  \hline
\ref{sec:detection} \textbf{Detection} in   \ref{sec:mythA},  \ref{sec:mythG}       & Visible resources and I/O & Existing methods  &   $\ominus$: ME is more visible by resources.    \\   \hline
\ref{sec:sysCall} \textbf{System calls} in  \ref{sec:mythH}     & Needs untrusted assistance   & Direct   access                    &   $\ominus$: ME has longer path to syscalls.    \\   \hline
\ref{sec:priviliges} \textbf{Privileges} in   \ref{sec:mythE},  \ref{sec:mythI}      & User level/ring 3  &  Kernel/ring 0          &   $\ominus$: ME is less privileged.    \\   \hline
\ref{sec:resources} \textbf{Resources}  in  \ref{sec:mythJ}       & Limited/restricted           & Unlimited                        &   $\ominus$: ME has access to less resources.    \\  
  \textbf{Access to user assets}  in  \ref{sec:mythD} &  Indirectly               &  Direct   Access                                &   $\ominus$: ME has longer path to user assets.    \\   \hline
\ref{sec:audience} \textbf{Audience } in  \ref{sec:mythK}           & CPU-specific                    & Independence                &   $\ominus$: ME has less number of targets.    \\   
 \textbf{Availability}  in  \ref{sec:mythC}       & Low  availability               & Always  On                                    &   $\ominus$: ME is less available.    \\   \hline
\ref{sec:maintenance} \textbf{Maintenance} in  \ref{sec:mythF}, \ref{sec:mythL}      & Specific dependencies  & Easy-to-resell      &   $\ominus$: ME is more expensive to maintain.    \\ \hline
    \end{tabular}
    \label{tab:compareMalware}
\end{table*}

\subsection{Enclave-Assisted Malware vs. Malware in the Wild}
\label{sec:comparison}
Table~\ref{tab:compareMalware} summarises the malware characteristics examined within the misconceptions presented in Section~\ref{sec:myths}.
An application from an outside, untrusted world must trigger the enclave-based malware (\ie~the enclave binary portion). 
Malware can only communicate with the outside world through untrusted channels.
The behaviour patterns of input and output streams are visible to detection mechanisms.
For system APIs, calls, lower-level support, an enclave needs assistance from the untrusted world.
Malware in an enclave runs at the user level privileges (ring 3 in monolithic systems).
System resources such as memory, threading, timers and CPU features are restricted and limited for enclaves.
The audience is limited to SGX-enabled (or TEE enabled) machines only.
Key management requires extra steps to manage the payload with PKI as usual, and enclave identity does not change based on the payload.
Enclave availability is lower than other system services; enclaves are not in constant or persistent operation.
Enclaves cannot directly access user assets, but must indirectly utilise outside resources for completing an attack.
Malware in the enclave requires specific dependencies in development and makes updates and reselling more challenging.

\subsection{Can SGX boost any characteristics of the malware?}

A wild malware in an untrusted system can place various attacks against enclaves.
In contrast, enclaves are limited and continue to be more limited in performing attacks on other enclaves and towards the untrusted world.
The design goal of enclaves is to protect the critical operations from wild malware.
Utilising an enclave for a malware weakens its abilities dramatically.

\section{Discussion on Malware and Trusted Execution Environments}
\label{sec:discussMalwareTEE}
TEE are introduced to limit the existing abilities of wild malware. 
Moving more applications into the TEE domain will eventually leave wild malware outside the trusted ecosystem.
Crucially, TCB minimisation plays a key role in leaving malicious software outside of the trusted applications.
Including a fully capable operating system together with the user assets inside an enclave may simply recreate the untrusted world and its problems again.

\subsection{Zero day SGX vulnerabilities in Malware as a Service (MaaS)}

SGX and TEE raised awareness of micro-architectural attacks.
Due to microcode weaknesses, an advanced malware can temporarily take advantage of vulnerable systems.
These zero-day vulnerabilities may be equipped in APTs.
However, TCB updates on microcode and revocation mechanisms make such attacks infeasible in the long term~\cite{intel_microcodeUpdate}.

To implement an advanced malware with SGX, attackers must utilise the vulnerabilities in the core SGX ecosystem defined in Section~\ref{sec:ecosystem}.
Intel can prevent such attacks by TCB updates.
If not patched, bad actors may sell the security problems of the core SGX ecosystem in exploit markets for MaaS.

\subsection{Potential malware planted inside SGX ecosystem}

In the unlikely event of a core SGX ecosystem including critical vulnerabilities, a sophisticated malware may be planted into benign enclaves (\eg~during compiling time, in a development environment, in SDKs, or trusted libraries).
Due to the increased use of hardware enclaves (\ie~moving of critical user assets into enclaves), attackers may target the core ecosystem more.
Future questions can be around malware in the core SGX ecosystem, not in a custom enclave but on a microcode level. 
\begin{itemize}
\item Are more attacks possible on micro-architectural level towards malware deployment, similar to attacks in ring minus two or three (ring -2, ring -3)~\cite{wojtczuk2011following,wojtczuk2012stitch,tereshkin2009introducing,wojtczuk2009attacking,wojtczuk2009attackingTXT_SINIT,wojtczuk2009another,wojtczuk2009attackingTXT,rutkowska2008preventing}, System Management Mode (SMM)?
\item Can malware authors infect/exploit the architectural enclaves with no one noticing?
\end{itemize}
After all, every software brings an attack vector; however, breaking the enclave's security may have a high impact as more critical assets move into enclaves.

\subsection{Malware capabilities in wild, without a TEE}

In contrast to SGX-based malware, a typical malware in wild, commodity and high noise systems, can utilise all of the system resources such as registry configurations, services, triggered events, drivers and other hardware devices for spreading and persistence.
Wild malware can hide its communication patterns in high noise, and distinguishing what belongs to malware behaviour remains an open challenge.
Malware detection techniques with memory analyses apply, but continue to be a challenge as wild malware can abuse any software package in an untrusted world.
Malicious applications in an untrusted world can often access system calls  directly as they are available to use and not distinguishable from benign use.
Most of the benign software already operates at the kernel level and malware in the wild continues to abuse any of the existing primitives.
System resources accessing main memory, and other CPU features are not limited for a wild malware.
Enclave independent malware can target any victim with a commodity computer (\eg~x86).
Malware authors utilise PKI and other crypto techniques in place in order to manage the master keys of their malware.
Availability of malware in the wild is up as long as the system operates as normal.
Malware in the wild operates in the same environment where user assets are placed, often giving direct access to valuable information.
Maintenance, updating, and reselling a malware in the wild can be straightforward by pulling another payload from a new malware author, or simply via passing/delegating keys to a buyer in a remote environment.

Malware with or without SGX can be detected with similar capabilities.
For example, in a cloud server, bad actors have the freedom to deploy malicious operations as a client of a cloud company.
Company admins may watch the network traffic for suspicious and criminal activities or due to complaints.
Normally, the attacker or the server tenant is free to run any software in the rented machine.
Using or not using SGX does not bring any difference to this model.
Alternatively, attackers may want to utilise enclaves for private operations such as domain name generation (to be used in a campaign), or for planning criminal activities.
Private operations, however, require an active connection and persistent communication to serve.
These are similar possibilities to be handled in remote servers and enclave utilisation does not bring any immediate benefit to these use cases.

\section{Conclusion}

We have examined why malware in a trusted execution environment will become much weaker than operating in a wild, high noise system.
We considered \emph{frequently seen characteristics} of an ideal malware.
With the use of TEE, these characteristics are either still the same as any typical malware in a system, or they are more restricted than before.
In the case studies, we revisited how a fully undetected and scalable malware infection campaign can already be in place, without any assistance of trusted hardware.
Malware distribution via social media platforms remains crucial compared to malware delivery through enclaves.
Section~\ref{sec:myths} showed the misconceptions about how enclaves operate and why relying on enclaves for adversarial purposes is a poor design choice.
In contrast, enclaves continue to be a powerful mechanism for protecting highly valuable user assets against malware.
Finally, in the near future, we may see more enclave-based malware, but they will be practically weaker than a malware in high entropy, and they will not be more superior than the malware samples in untrusted environments.
Known techniques and methods must be applied to eliminate these attempts.
We highlight that the communication and key management characteristics of the malware are seen in four myths in our paper.
We conclude that for these characteristics, the use of TEE makes the malware weaker in commodity systems.
Systematisation, categorisation and the definition of the myths can be done in many ways.
In future work, the myths and the characteristics can be extended and studied in more depth to see whether the arguments hold for future TEE or not. 
Considering the new hardware prototypes supporting enclave-like isolated containers, our structured discussions on malware can help mitigate potential TEE abuse at the design stage.


\section*{Acknowledgements}
We thank anonymous reviewers for their feedback.
This work is partly supported by a grant from InnovateUK under grant number 105592.
We thank Raoul Strackx, A Acar, David Grawrock, Michał Kowalczyk and anonymous colleagues for their feedback and discussions.
 

\bibliographystyle{ACM-Reference-Format}
\bibliography{refs}

\appendix

\end{document}